\documentclass[12pt]{article}
\input epsf
\usepackage{amssymb}
\usepackage{latexsym}

\def \be{\begin{equation}}
\def \ee{\end{equation}}
\def \berr{\begin{eqnarray}}
\def \err{\end{eqnarray}}
\def \nn{\nonumber}
\def \a{\alpha}

\def \b{\beta}

\def \Del{\Delta}

\def \L{\Lambda}
\def \Om{\Omega}

\def \eps{\varepsilon}

\def \om{\omega}

\def \A{{\cal A}}

\def \RR{{\cal R}}

\def \Tr{\mbox{Tr}}
\def \S{S^2}

\def \({\left(}
\def \){\right)}
\def \<{\langle}
\def \>{\rangle}
\def \[{\left[}
\def \]{\right]}

\def\tens{\mathop{\otimes}}

\newcommand{\tr}{\triangleright}

\def\reps{representations }
\def\rep{representation }

\def\mf#1{{\mathbb #1}}

\def\R{{\mf{R}}}
\def\C{{\mf{C}}}

\newcommand{\sect}[1]{\setcounter{equation}{0}\section{#1}}

\evensidemargin \oddsidemargin
\begin{document}


\begin{center}  

{\Large\bf
Aspects of the $q$--deformed Fuzzy Sphere\footnote{Talk presented at 
the Euroconference ``Brane New World and Noncommutative Geometry''
in Villa Gualino, Torino, Italy, October 2 - 7, 2000.}
 \\[2ex]}
Harold \ Steinacker
\\[2ex] 
{\small\it 
        Laboratoire de Physique Th\'eorique et Hautes Energies\\
        Universit\'e de Paris-Sud, B\^atiment 211, F-91405 Orsay \\[2ex] }
\end{center}

\begin{abstract}
These notes are a short review of the $q$--deformed 
fuzzy sphere $S^2_{q,N}$, which is a ``finite'' noncommutative 2--sphere 
covariant under the quantum group $U_q(su(2))$. 
We discuss its real structure, differential calculus and integration
for both real $q$ and $q$ a phase, and show how actions for
Yang--Mills and Chern--Simons--like gauge theories arise naturally.
It is related to $D$-branes on the $SU(2)_k$ WZW model for 
$q = \exp(\frac{i \pi}{k+2})$.
\end{abstract}

\sect{Introduction}

$S^2_{q,N}$ is a $q$--deformed version of the ``ordinary'' fuzzy sphere
$S_N^2$ \cite{madore}.
It is covariant under the standard Drinfeld--Jimbo quantum group
$U_q(su(2))$, and can be defined for both $q \in \R$ and $|q| = 1$. 
The algebra of functions on $S^2_{q,N}$ is isomorphic to 
the matrix algebra $Mat(N+1,\C)$, but it carries additional structure which 
distinguishes it from $S_N^2$, related to its rotation symmetry under
$U_q(su(2))$. 
For real $q$, we recover precisely the ``discrete series'' of Podles spheres 
\cite{podles}. We describe its structure in general,
including a covariant differential calculus
and integration, and show how actions of Yang--Mills and Chern--Simons
type arise naturally on this space. 
A much more detailed study of $S^2_{q,N}$ has been given by in 
\cite{fuzzysphere}. These considerations were 
motivated mainly by the work \cite{ARS} of Alekseev, 
Recknagel and Schomerus, who study the boundary conformal field
theory describing spherical $D$--branes in the $SU(2)$ WZW model at 
level $k$. These authors extract an ``effective'' algebra of functions on the
$D$--branes from the OPE of the boundary vertex operators.
This algebra is twist--equivalent \cite{fuzzysphere}
to the space of functions on $S^2_{q,N}$, if 
$q$ is related to the level $k$ by the formula
\be
q=\exp(\frac{i \pi}{k+2}).
\label{q}
\ee

\sect{The space $S_q^2$}

First, recall that an algebra $\A$ is a
$U_q(su(2))$--module algebra if there exists a map 
\berr
U_q(su(2)) \times \A &\rightarrow& \A, \nn\\
  (u,a) &\mapsto& u \tr a
\err
which satisfies $u \tr (ab) = (u_{(1)} \tr a) (u_{(2)} \tr b)$
for $a, b \in \A$. Here $\Delta(u) = u_{(1)} \tens u_{(2)}$
is the Sweedler notation for the coproduct of $u \in U_q(su(2))$. 

A particularly simple way to define the $q$--deformed fuzzy sphere
is as follows: Consider the spin $\frac N2$ \rep of $U_q(su(2))$, 
\be
\rho: U_q(su(2)) \rightarrow Mat(N+1,\C), 
\label{rho}
\ee
which acts on $\C^{N+1}$. With this in mind, it is natural to consider
the simple matrix algebra $Mat(N+1,\C)$
as a $U_q(su(2))$--module algebra, by
$u \tr M = \rho(u_1) M \rho(Su_2)$. This defines $S^2_{q,N}$.
It is easy to see that 
under this action of $U_q(su(2))$, it decomposes into the irreducible \reps 
\be
S^2_{q,N}:= Mat(N+1,\C) = (1) \oplus (3) \oplus ... \oplus (2N+1)
\ee 
(if $q$ is a root of unity (\ref{q}), this holds 
provided $N \leq k/2$, which we will assume here). 
Let $\{x_i\}_{i = +,-,0}$ be the weight basis of the spin 1 components,
so that $u \tr x_i = x_j \pi^j_i(u)$ for $u \in U_q(su(2))$.
One can then show that they satisfy the relations 
\berr
\eps^{ij}_k x_i x_j &=& \L_{N} x_k,  \nn\\
g^{ij} x_i x_j  &=& R^2. 
\err
Here 
\be
\L_{N} = R\; \frac{[2]_{q^{N+1}}}{\sqrt{[N]_q [N+2]_q}},
\ee
$[n]_q = \frac {q^n-q^{-n}}{q-q^{-1}}$, and 
$\eps^{ij}_k$ and $g^{ij}$ are the $q$--deformed 
invariant tensors. For example, $\eps^{33}_3 = q^{-1} - q$, and 
$g^{1 -1} = -q^{-1}, \;\; g^{0 0} = 1, \;\; g^{-1 1} = -q$. 
In \cite{fuzzysphere}, these relations 
were derived using a Jordan--Wigner construction. 
For $q=1$, the relations of $S^2_N$ are recovered, and for real
$q\neq 1$  we obtain 
$\eps^{ij}_k x_i x_j = R\;(q-q^{-1})\; x_k$ in the limit 
$N \rightarrow \infty$.

\paragraph{Real structure.}
In order to define a {\em real} noncommutative space, one must 
specify a star structure on the algebra of functions. Since 
$S^2_{q,N}$ should decompose into unitary \reps of $U_q(su(2))$,
we restrict to the cases $q\in \R$  and  $|q| = 1$. 
In either case, the star structure on $S^2_{q,N} = Mat(N+1,\R)$
is defined to be the usual hermitean adjoint of matrices. 
In terms of the generators  $x_i$, this  becomes 
\be
x_i^* = g^{ij} x_j, \qquad\mbox{if}\; q \in \R
\ee 
and 
\be
x_i^* = - \om x_i \om^{-1} = x_j \rho({L^-}^{j}_{k}) q^{-2} g^{k i}, 
       \qquad\mbox{if}\; |q| = 1.
\ee
Here $\om  \in \hat U_q(su(2))$ generates the quantum Weyl reflection
\cite{kirill_resh}, 
\be
\Del(\om) = \RR^{-1}\om\tens \om,    \qquad
\om^2 = v \epsilon, \qquad v = S \RR_2 \RR_1 q^{-H},
\ee
and
${L^-}^{i}_{j} = \pi^{i}_{j}(\RR_1^{-1}) \RR_2^{-1} \in U_q(su(2))$.
Using the map $\rho$ (\ref{rho}), this amounts to the star structure
$H^* = H, (X^{\pm})^* = X^{\mp}$ defining the compact form $U_q(su(2))$
for both $q\in \R$  and  $|q| = 1$. 
In the case $q \in \R$, we have recovered precisely the discrete series of 
Podles spheres \cite{podles}. 

To summarize, $S^2_{q,N}$ is same {\em algebra} $Mat(N+1,\C)$ as 
$S^2_N \equiv S^2_{q=1,N}$, but its symmetry $U_q(su(2))$ acts on it in 
a way which is inequivalent to the undeformed case.
It admits additional structure compatible with this symmetry,
such as a differential calculus and an integral. This will be discussed 
next.

\sect{Differential calculus and integration}

In order to write down Lagrangians, it is convenient to use the notion 
of an (exterior) differential calculus. 
A covariant differential calculus over $S_{q,N}$ is a 
graded bimodule $\Omega^* = \oplus_n \; \Omega^n$
over $S_{q,N}$ which is a $U_q(su(2))$--module algebra, 
together with an exterior derivative $d$ which satisfies $d^2=0$ and 
the graded Leibnitz rule.

The structure of the calculus is determined by requiring covariance, and 
a systematic way to derive it is given in \cite{fuzzysphere}. 
Here we will simply quote the most important features.
First, the modules $\Omega^n$ turn out to be free over $S_{q,N}$
both as left and right modules\footnote{but not as bimodules}
with $dim \;\Om^n = (1,3,3,1)$ for $n = (0,1,2,3)$, and vanish for higher $n$. 
In particular, it is not possible to have a calculus with only ``tangential''
forms; this means that vector fields over $S_{q,N}$ will in general
also contain ``radial'' components. As suggested by the dimensions,
there exists a canonical map
\be
\ast_H: \; \Omega^n \rightarrow \Omega^{3-n},
\ee
which satisfies $(\ast_H)^2 = id$, and
respects the $U_q(su(2))$-- and $S_{q,N}$--module structures. It satisfies
furthermore
$$
\a (\ast_H \b) = (\ast_H \a) \b
$$
for any $\a, \b \in \Omega^*$.
Moreover, there exists a special one--form
$$
\Theta \in \Om^1_{q,N}
$$
which is a singlet under $U_q(su(2))$, and generates the calculus 
as follows:
\berr
d f &=& [\Theta,f],  \nn\\
d\a^{(1)} &=& [\Theta,\a^{(1)}]_+ - \ast_H(\a^{(1)}),  \label{d_1}\\
d\a^{(2)} &=& [\Theta,\a^{(2)}]  \nn
\err
for any $f \in S_{q,N}$ and $\a^{(i)}\in\Omega^i$.
One can verify that
$$
d\Theta = \Theta^2 = \ast_H(\Theta),
$$
and $[f, \Theta^3] = 0$ with $\Theta^3 \neq 0$.
There also exists a star structure \cite{fuzzysphere}
on $\Omega^*$ for 
both $q \in \R$ and $|q| = 1$, which makes it a covariant $\star$ calculus.

\paragraph{Frame.}
The most convenient basis to work with is the ``frame'' generated by 
one--forms $\theta^a\in\Omega^1$ for $a = +,-,0$, which satisfy
\berr
[\theta^a, f] &=& 0,        \\
\theta^a \theta^b  &=& -q^2 \;\hat R^{ba}_{cd}\theta^d \theta^c,  \\
\ast_H \theta^a    &=& - \frac 1{q^{2}[2]_{q^2}} 
                       \eps^a_{bc} \theta^c \theta^b, \nn\\
\theta^a \theta^b \theta^c &=& -\L_N^2 \frac{q^6}{R^2} \;\eps^{cba} \Theta^3.
\err
Such $\theta^a$ exist and are essentially unique. The disadvantage is that 
they have a somewhat complicated transformation law
$$
u \tr \theta^a = u_1 Su_3 \pi^a_b(Su_2)\; \theta^b.
$$
One can alternatively use a basis of one--forms $\xi_i$ which transform 
as a vector under $U_q(su(2))$, but then the commutation relations 
are more complicated \cite{fuzzysphere}.
\paragraph{Integration.}
The unique invariant integral of a function $f \in S^2_{q,N}$ is 
given by its quantum trace as an element of $Mat(N+1,\C)$,
$$
\int f :=  \frac {4\pi R^2}{[N+1]_q} \Tr_q(f) = 
           \frac {4\pi R^2}{[N+1]_q} \Tr(f \;q^{-H}),
$$
normalized such that $\int 1 =4\pi R^2$.
Invariance means that $\int u \tr f = \eps(u) \; \int f$.
It is useful to define also the integral of forms, by declaring
$\Theta^3$ to be the ``volume form''. Writing any 3--form as 
$\a^{(3)} = f \Theta^3$, 
we define
$\int \a^{(3)} = \int f \Theta^3 := \int f$.
Using the correct cyclic property of this integral, one can then verify  
Stokes theorem 
\be
\int d\a^{(2)} =  \int [\Theta,\a^{(2)}] = 0.
\ee

\sect{Gauge Fields}
Actions for gauge theories arise in a very natural way on $\S_{q,N}$. 
We shall describe (abelian) gauge fields as one--forms
$$
B = \sum B_a \theta^a  \quad \in \Omega^1,
$$ 
expanded in terms of the frames $\theta^a$.
They have 3 independent components, which
means that $B$ has also a radial component. 
However, the latter cannot be disentangled from the other components, 
since it is not possible to construct a covariant 
calulus with 2 tangential components only.

We propose that Lagrangians for gauge fields should contain {\em no}
explicit derivatives in the $B$ fields. The kinetic terms will then arise 
naturally upon a shift of the field $B$,
$$
B = \Theta + A.
$$
The only Lagrangian of order $\leq 3$ in $B$ which after this shift
contains no linear terms in $A$ is the `` Chern-Simons'' action
\be
S_{CS} := \frac 13 \int B^3 - \frac 12 \int B \ast_H B = - \mbox{const}
    + \frac 12  \int A dA + \frac 23 A^3.
\label{S_CS}
\ee
Going to order 4 in $B$, we define the curvature as 
$$
F := B^2 - \ast_H B = dA + A^2,
$$
using (\ref{d_1}). Then a ``Yang--Mills'' action is naturally obtained as
\be 
S_{YM} :=  \int F \ast_H F =  \int (dA + A^2) \ast_H (dA + A^2).
\label{YM}
\ee
These are precisely the kind of actions that have been found \cite{ARS}
in the string--induced low--energy effective action on $D$--branes
in the $SU(2)_k$ WZW model, in the leading approximation.
\paragraph{Acknowledgements}
The author is very grateful for the collaboration with H. Grosse and
J. Madore on the paper \cite{fuzzysphere} which underlies this review,
and to the organizers of the Euroconference 
``Brane New World and Noncommutative Geometry'' for the invitation and 
support.

\end{document}